\documentclass[prb,superscriptaddress,showpacs,twocolumn]{revtex4}%
\usepackage[english]{babel}

\usepackage{amsmath,amssymb,braket,mathtools,array,MnSymbol,bm,hhline}
\usepackage{graphicx,color}
\usepackage{braket}

\usepackage[thinlines]{easytable}

\graphicspath{{images/}}

\usepackage{hyperref}
\hypersetup{
	colorlinks   = true, 
	urlcolor     = blue, 
	linkcolor    = blue, 
	citecolor   = blue 
}

\newcommand{\abs}[1]{\left|#1\right|}

\newcommand{\appropto}{\mathrel{\vcenter{
			\offinterlineskip\halign{\hfil$##$\cr
				\propto\cr\noalign{\kern2pt}\sim\cr\noalign{\kern-2pt}}}}}

\renewcommand{\vec}[1]{{\boldsymbol #1}}

\newcommand{\ucheck}[1]{{\check{\underline #1}}}

\renewcommand{\Im}{{\rm \, Im\,}}

\newcommand{\rmi}{\mathrm{i}}
\newcommand{\Tr}{{\rm Tr\,}}
\newcommand{\sgn}{{\rm sgn\,}}
\renewcommand{\d}{\mathrm{d}}

\renewcommand{\S}{{\cal S}}

\begin{document}
	
\title{Superfluid density of a photo-induced superconducting state}

\author{A. Shtyk}
\affiliation{Department of Physics, Harvard University, Cambridge, MA 02138, USA}

\author{G. Goldstein}
\affiliation{Cavendish Laboratory, University of Cambridge, Cambridge, CB3 0HE, United Kingdom}

\author{C. Aron}
\affiliation{Laboratoire de Physique Th\'eorique, \'Ecole Normale Sup\'erieure, \\
CNRS, PSL University, Sorbonne Universit\'e, Paris 75005,
France}
\affiliation{Instituut voor Theoretische Fysica, KU Leuven, Belgium}
 
\author{C. Chamon}
\affiliation{Department of Physics, Boston University, Boston, MA 02215, USA}

\begin{abstract}
Nonequilibrium conditions offer novel routes to superconductivity that
are not available at equilibrium.  For example, by engineering
nonequilibrium electronic populations, pairing may develop between
electrons in different energy bands.  A concrete proposal has been
made to photo-induce superconductivity in a semiconductor, where
pairing occurs between electrons in the conduction and valence bands,
even for repulsive interactions. Here, we calculate the superfluid
density for such a nonequilibrium paired state, and find it to be
positive for repulsive interactions and interband pairing. The
positivity of the superfluid density implies the stability of the
photo-induced superconducting state as well as the existence of the
Meissner effect.

\end{abstract}

\maketitle

\section{Introduction}

The subject of nonequilibrium superconductivity has recently gained
much interest, in part due to experiments on photo-induced transient
states in YBa$_2$Cu$_2$O$_{6+\delta}$\cite{kaiser2014}, and the
subsequent experimental and theoretical progress (see
Ref.~\onlinecite{Review_Cavalleri} for a review). In fact, ideas to
extend the temperature regime where superconductivity exists by
optical excitation have a long history. Their root can be traced back
to the pioneering theoretical predictions by Eliashberg and
Gor'kov\cite{gorkov1968,eliashberg1970}, who showed that microwave
radiation may assist the formation of the superconducting gap and thus
raise the transition temperature $T_\mathrm{c}$. These predictions
were later confirmed by experiments\cite{dmitriev1970}. The applied
electromagnetic radiation shifts the electronic occupation numbers,
extending the temperature regime in which the BCS self-consistency
equation has a non-zero solution.

These ideas of population control can also be applied to systems that
are not superconductors at equilibrium. It was proposed that
superconductivity could be induced in narrow, indirect gap
semiconductors, with pairing between electrons in the same band
(intraband
pairing)\cite{galitskii1973,kirzhnits1973,elesin1973theory,galitskii1986}. A
non-zero superconducting gap was shown to be possible with attractive
and, notably, with repulsive electronic interactions as well. The
latter case is particularly important because repulsion is prevalent
in electronic systems. However, in the latter case it was also noted
that there was no Meissner effect accompanying the formation of a gap,
because the sign of the current response was opposite to that in a
standard superconductor: the system would respond as a perfect
paramagnet instead of a perfect diamagnet. This strange response,
corresponding to a \textit{negative} superfluid density, signals that
the state is unstable for repulsive interactions.

Recently, Ref.~\onlinecite{goldstein2015} proposed to use optical
pumping to achieve \emph{interband}, rather than intraband,
pairing. In this scheme, electrons sitting in two bands at widely
different energies and far away from the chemical potential can form
interband Cooper pairs in the $s$-wave channel, even in the case of
repulsive interactions. (See Figure 1 for a sketch of the relevant
mechanism.)

In this paper, we investigate the stability of the corresponding 
photo-induced interband superconducting state by computing its superfluid
density. We find a \emph{positive} superfluid density for repulsive
interactions for all parameters (\textit{e.g.}, band curvatures,
quasi-particle populations) where a non-trivial mean-field solution of
the self-consistent BCS equation exist. The positivity of the
superfluid density implies the stability of the state as well as the
existence of a Meissner effect. The later could be used as a reliable
alternative to transport properties to confirm the presence of a
nonequilibrium superconducting order in a semiconductor.

\begin{figure}[t!]
\center{\includegraphics[width=.71\linewidth]{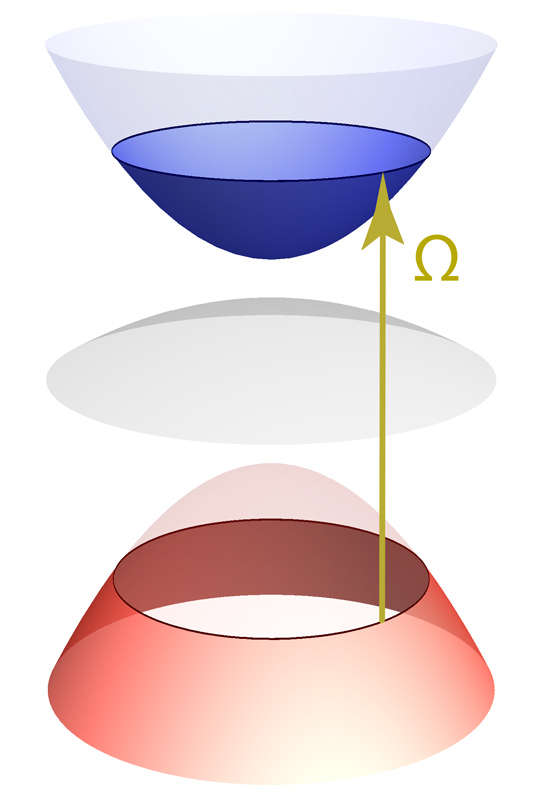}} 
\caption{(color online) Electrons of a semiconductor are optically
  pumped from the valence (red) band to the conduction band (blue). In
  the absence of pumping, \textit{i.e.} in thermal equilibrium, the
  conduction band is empty. The laser drive creates a nontrivial
  steady-state distribution function partially depleting (populating)
  the valence (conduction) band. Such a population imbalance can also
  be assisted \textit{via} an auxiliary band (gray).  }
	\label{fig:1}
\end{figure}

\section{Model}

We consider a two-band semiconductor model with electronic dispersions $E_{1\vec{p}}$ and $E_{2\vec{p}}$. The chemical potential of the system is set in the middle of the two bands, see Fig.~\ref{fig:1}. For simplicity, we consider symmetric bands, \textit{i.e.} $E_{\alpha \, \vec{p}} = E_{\alpha \, -\vec{p}}$ for $\alpha = 1, 2$.
The semiconductor is optically pumped with a broad-band light source, as described in Ref.~\onlinecite{goldstein2015}. In this setup, the optical pumping creates a nonequilibrium distribution of the quasiparticles, which is key to  achieve the interband pairing.

\subsection{Hamiltonian}
Let us assume that, after a transient regime, the interband pairing has already built up. The mean-field description of the system consists in the following BCS Hamiltonian
\begin{align} \label{eq:H1}
H=\sumint_{\vec{p}\alpha} \!\!  E_{\alpha \vec{p}}c_{\vec{p}}^{\alpha\dagger} c_{\vec{p}}^\alpha
+\frac{1}{2}\sumint_{\vec{p}\alpha\beta}\Big[c_{\vec{p}}^{\alpha\dagger} \Delta(i\sigma^y_{\alpha\beta})c_{-\vec{p}}^{\beta\dagger} + \text{h.c.}\Big],
\end{align}
where $c_{\vec{p}}^{\alpha\dagger}$ and $c_{\vec{p}}^\alpha$ are the fermionic creation and annihilation operators corresponding the electrons in the conduction ($\alpha = 1$) and  valence ($\alpha=2$) bands. To simplify the discussion, we work with spinless electrons. 
$\Delta$ is the $s$-wave superconducting order parameter to be determined self-consistently, see below. $\sigma^y$ is the usual Pauli matrix. Without loss of generality, we assume that $\Delta$ is real. 
After introducing the Nambu-Gor'kov spinor notation
\begin{align}
\label{NG}
	\Psi_\vec{p}=
	\begin{pmatrix}
	c_{\vec{p}}^1\,,& c_{-\vec{p}}^{2\dagger}
	\end{pmatrix}\,,
\end{align}
the Hamiltonian in Eq.~(\ref{eq:H1}) can be compactly re-written as
\begin{align} \label{eq:H2}
	H=\int_{\vec{k}}\Psi^{\dagger}_\vec{p}[\varepsilon_{\vec{p}} \check{I} +E_{\vec{p}}\check{\tau}^z+\Delta\check{\tau}^x] \Psi_\vec{p},
\end{align}
where
\begin{align}
E_{\vec{p}} \equiv (E_{1 \vec{p}}+E_{2 \vec{p}})/2 \, ,\\
\varepsilon_{\vec{p}} \equiv (E_{1 \vec{p}}-E_{2 \vec{p}})/2 \,,
\end{align}
and where $\check{I}$, $\check{\tau}^x$, and $\check{\tau}^z$ are the usual identity and Pauli matrices, respectively, in Nambu-Gor'kov space.
 
The Hamiltonian can be readily diagonalized \textit{via} the following
transformations in Nambu-Gor'kov space
\begin{align}
\label{U}
\check{U}_\vec{p} = \exp \left[ \frac{\rmi}{2}\beta_\vec{p}\check{\tau}_y \right]\,, \mbox{ with } \tan\beta_\vec{p} \equiv \frac{\Delta}{E_\vec{p}} \,, 
\end{align}
leading to
\begin{align}
\underline{H}
 =  \int_{\vec{p}} \Psi^{\dagger}_\vec{p} [ \xi_\vec{p} \ucheck{\tau}^z+\varepsilon_\vec{p} \ucheck{I} ]  \Psi_\vec{p} \,,
\end{align}
with $\xi_\vec{p} \equiv \sqrt{E_\vec{p}^2+\Delta^2}$. We adopt the underlining to indicate the use of the quasiparticle basis.

\subsection{Keldysh action}
We compute the electrodynamic response of a generic steady state within the Schwinger-Keldysh formalism~\cite{keldysh,Rammer,Kamenev2009}. The Keldysh action corresponding to the Hamiltonian in Eq.~(\ref{eq:H2}) reads
\begin{align}
\S= \int_\Upsilon \!\! \d t \sumint_{\vec{p}} \Psi^{\dagger}_\vec{p} [  \left( \rmi\partial_t-\varepsilon_{\vec{p}} \right) \check{I} -E_{\vec{p}} \check{\tau}^z-\Delta\check{\tau}^x ] \Psi_\vec{p} \,,
\end{align}
where the time integral over $t$ goes along the standard Keldysh contour $\Upsilon$ going from $-\infty$ to $+\infty$ and then back to $-\infty$. It is customary to perform a so-called Keldysh rotation of the fields living on those two time branches. In the resulting $2 \times 2$ Keldysh space, the electron Green's functions are organized as
\begin{align}
\check{\boldsymbol{G}}=
\begin{pmatrix}
	\check{G}^R & \check{G}^K
	\\
	0 & \check{G}^A
\end{pmatrix},
\end{align}
where $\check{G}^R$,  $\check{G}^A$, and  $\check{G}^K$ are the retarded, advanced and Keldysh Green's functions, respectively.
The retarded Green's function encodes the spectral properties of the steady-state and reads $\check{G}^R=\check{U} \ucheck{G}^R \check{U}^\dagger$ with 
\begin{align}
\label{eq:G}
\ucheck{G}^R(\epsilon,\vec{p})^{-1}
=  \left( \epsilon -\varepsilon_{\vec{p}} + \rmi 0 \right) \ucheck{I} - \xi_\vec{p} \ucheck{\tau}^z,
\end{align}
where the operator $\check{U}$ was defined in Eq.~(\ref{U}). A convenient representation is given by 
\begin{align}
\check{G}^R(\epsilon,\vec{p})=&\frac{1}{  \epsilon-\varepsilon_\vec{p} - \xi_\vec{p} +\rmi 0} \check{P}_{+ \vec{p}} +\frac{1}{\epsilon-\varepsilon_\vec{p}+ \xi_\vec{p}+\rmi 0}\check{P}_{- \vec{p}}
\end{align}
with $\check{P}_\pm$ being the projectors onto the states of the
quasiparticle basis, \textit{i.e.} $\ucheck{P}_\pm=(\ucheck{I}
\pm\ucheck{\tau}^z)/2$. $\check{G}^A$ can be determined from
$\check{G}^R$ by a simple time-reversal operation. The Keldysh Green's
function encodes the nonequilibrium state populations,
\begin{align}
\label{eq:K}
	\check{G}^K(\epsilon,\vec{p})=\left[\check{G}^R\check{F}-\check{F}\check{G}^A\right]{(\epsilon,\vec{p})},
\end{align}
with the matrix $\hat{F}(\epsilon,\vec{p})$ encoding a general
nonequilibrium electron distribution function. In general, $\check{F}$
does not commute with the Green's function $\check{G}^{R(A)}$, but in
thermal equilibrium the situation is greatly simplified and
$\check{F}(\epsilon,\vec{p})$ becomes
$\check{F}(\epsilon,\vec{p})=\tanh\epsilon/2T\cdot \ucheck{I}$.

The electromagnetic vector potential $\vec{A}$ generates an electric current $\boldsymbol{J}$ whose spatial components $J^\mu$ ($\mu = x, y , z$) read
\begin{align}
\label{J}
	J^\mu=e\sumint_{\vec{p} \nu}  \langle \Psi^{\dagger}_\vec{p} \left[\hat{v}^\mu_{\vec{p}}-e\hat{\partial_\mu v^\nu_{\vec{p}}} \, A^\nu\right] \Psi_\vec{p} \rangle,
\end{align}
where $ e < 0$ is the charge of the electron and we set the speed of light $c=1$, while the ``velocity'' and ``mass'' are
\begin{align}
	\check{v}^\mu_\vec{p}=V^\mu_\vec{p} \check{I} +v^\mu_\vec{p} \check{\tau}^z,\quad
	\check{\partial_\mu v^\nu_\vec{p}}=\partial_\mu V^\nu_\vec{p}\check{\tau}^z+\partial_\mu v^\nu_\vec{p} \check{I},
\end{align}
with $\partial_\mu \equiv \partial / \partial p^\mu$,  $V_\vec{p}^\mu \equiv \partial_\mu E_\vec{p}$, and $v_\vec{p}^\mu \equiv \partial_\mu \varepsilon_\vec{p}$. In the quasiparticle basis, they read (omitting the $\vec{p}$ indices)
\begin{align}
\begin{split}
\ucheck{v}^\mu = &V^\mu \ucheck{I} +v^\mu(\ucheck{\tau}^z \cos\beta - \ucheck{\tau}^x \sin\beta ) \,,
\\
\label{eq:pv}
\underline{\check{\partial_\mu v^\nu}}=& \partial_\mu V^\nu (\ucheck{\tau}^z\cos\beta-\ucheck{\tau}^x\sin\beta) +\partial_\mu  v^\nu \ucheck{I} \,.
\end{split}
\end{align}

\subsection{Self-consistency equation}

In this manuscript,  we consider a generic steady state with a diagonal quasiparticle distribution function,
\begin{align}
\label{eq:n}
	\Braket{\Psi_\vec{p}^{\alpha\dagger}\Psi_\vec{p}^\beta}=
	\check{U}^\dagger\begin{pmatrix}
	\underline{n}_{1 \vec{p}} & 0
	\\
	0 & 1- \underline{n}_{2\vec{p}}
	\end{pmatrix}\check{U} \,.
\end{align}
The equation (\ref{eq:n}) implies that the matrix $\check{F}$ encoding the quasiparticle distribution in Eq.~(\ref{eq:K}) is also diagonal in quasiparticle basis and is given by
\begin{equation}
	\ucheck{F}(\epsilon,\vec{p})=
	\begin{pmatrix}
	1-2\underline{n}_{1 \vec{p}} & 0
	\\
	0 & 2\underline{n}_{2\vec{p}}-1
	\end{pmatrix}.
\end{equation}
Moreover, this means that $\underline{\check{F}}$ commutes with the
Green's functions $\ucheck{G}^{R}$ and $\ucheck{G}^{A}$. To derive the
self-consistency equation, we go back to the original microscopic
electron interaction and track the origin of the superconducting
pairing in Eq.~(\ref{eq:H1}) as stemming from an electronic
interaction
\begin{align}
H_{\rm int}=V_{\rm int}\int_{\vec{p}_1, \vec{p}_2} \!\! c_{\vec{p}_2}^{2\dagger} c_{\vec{p}_1}^{2} c_{-\vec{p}_2}^{1\dagger} c_{-\vec{p}_1}^{1},
\end{align}
where $V_{\rm int}$ is positive for repulsive interactions. Within a mean-field treatment, we have
\begin{align}
\begin{split}
H_{\rm int}& =-\int_{\vec{p}}\left(\Delta c_{\vec{p}}^{2\dagger} c_{-\vec{p}}^{1\dagger}
+\Delta^*c_{\vec{p}}^{2}c_{-\vec{p}}^{1}\right),
\end{split}
\end{align}
together with a self-consistency equation reading
\begin{align}
\begin{split}
\Delta=&V_{\rm int}\int_{\vec{p}}\Braket{c_{\vec{p}}^{2}c_{-\vec{p}}^{1}}.
\end{split}
\end{align}
Using the explicit form of the distribution function in Eq.~(\ref{eq:n}), the self-consistency equation takes a standard form when written in terms of quasiparticle distribution function:
\begin{align}
\label{SCE}
1=\frac{V_{\rm int}}{2}\int_{\vec{p}}\frac{\underline{n}_{ 1 \vec{p}}+\underline{n}_{ 2 \vec{p}}-1}{\xi_\vec{p}},
\end{align}
where $\xi_\vec{p} \equiv \sqrt{E_\vec{p}^2+\Delta^2}$.
The nonequilibrium effects enter through changes of the quasiparticle distribution functions with respect to their equilibrium values,
\begin{align}
\underline{n}^{\rm eq}_{1 (2) \vec{p}}= n_{\rm F}(\varepsilon_\vec{p} \pm \xi_\vec{p}, \mu) \,,
\end{align}
where $n_{\rm F}(\epsilon,\mu) = [1+\exp(-(\epsilon-\mu)/T)]^{-1}$ is the Fermi-Dirac distribution at the temperature $T$ and chemical potential $\mu$. The latter would reproduce the standard BCS self-consistency equation.

\section{Superfluid density}
The focus of this paper is on the Meissner effect \textit{i.e.} the response of the system to a non-uniform static electromagnetic vector potential. In this static limit, the electromagnetic properties are governed by the superfluid density defined through the following relation
\begin{align}
J^\mu=-\rho^{\mu\nu} A^\nu.
\end{align}
We consider the isotropic case, when the superfluid density tensor $\rho^{\mu\nu}$ is reduced to a scalar quantity $\rho$,
\begin{equation}
\rho^{\mu\nu}=(e^2/d)\rho \, \delta_{\mu\nu}.
\end{equation}
The superfluid density $\rho$ in the present text differs from the standard  definition by a factor $e^2/d$, where $d$ is the spatial dimension, in order to simplify the expressions below.

Similarly to the equilibrium case, there are two contributions to the electric current, paramagnetic and diamagnetic, which stem respectively from the first and second term of the current expression in Eq.~(\ref{J}),
\begin{align}
\rho=\rho_{(\text{para})} + \rho_{(\text{dia})} \,.
\end{align}
The paramagnetic term is also often called gradient term, for in the
case of the parabolic band with electron mass $m$ we have
$\vec{v}=\vec{p}/m=(-\rmi\hbar/m)\boldsymbol{\nabla}$. Below, we compute
these two contributions for generic quasiparticle distributions, and
later specialize to the case of photo-induced superconductivity.

\subsection{Diamagnetic contribution}

We start with the diamagnetic contribution stemming  from the second term in Eq.~(\ref{J}),
\begin{align}
\label{eq:dia_1}
J^{\mu}_{(\text{dia})}=-e^2\sumint_{\vec{p}, \nu}
\Braket{ \Psi^{\dagger}_\vec{p} \left(\check{\partial_\mu v^\nu} A^\nu\right)\Psi_\vec{p} }.
\end{align}
This yields the superfluid density 
\begin{align}
\label{eq:dia_2}
\rho_{(\text{dia})}
=
-\frac{\rmi}{2}\sumint_{\epsilon,\vec{p},\mu}
\Tr\left[ \partial_\mu \check{v}_\mu \left[ \check{G}^K-\check{G}^R+\check{G}^A \right] (\epsilon,\vec{p}) \right]\,,
\end{align}
where the trace runs in Nambu-Gor'kov space. 
%
The equation~(\ref{eq:dia_2}) has contributions from both quasiparticle bands. The contribution from the upper quasiparticle band can be computed 
using equations~(\ref{eq:G}), (\ref{eq:pv}), and~(\ref{eq:n}), yielding
\begin{align}
\sumint_{\vec{p},\mu}
 \left(\partial_\mu V^\mu \cos\beta+\partial_\mu v^\mu \right) \underline{n}_{1 \vec{p}},
\end{align}
since the actual occupation number~\cite{Rammer} is given by
\begin{align}
	\underline{n}_{ 1 \vec{p}}=-\frac{\rmi}{2}\int_\epsilon  \left[\ucheck{G}_{11}^K-\ucheck{G}_{11}^R+ \ucheck{G}_{11}^A\right]{(\epsilon,\vec{p})}.
\end{align}
The second quasiparticle band contributes with
\begin{align}
\sumint_{\vec{p}, \mu}
\left(-\partial_\mu V^\mu \cos\beta_\vec{p}+\partial_\mu v^\mu \right) (1-\underline{n}_{2 \vec{p}}) \,,
\end{align}
such that, at last, we obtain
\begin{align}
\label{dia}
\begin{split}
\rho_{(\text{dia})}
=\sumint_{\vec{p},\mu} (\partial_\mu V^\mu \cos\beta_\vec{p} \, (\underline{n}_{ 1 \vec{p}}+\underline{n}_{2 \vec{p}}-1)+
\\
+(\partial_\mu v^\mu)(\underline{n}_{ 1 \vec{p}}-\underline{n}_{ 2 \vec{p}}+1) \, .
\end{split}
\end{align}

\subsection{Paramagnetic contribution}

The paramagnetic contribution stems from the first term of the current expression in Eq.~(\ref{J}), and it reduces to the current-current correlation function,
\begin{align}
J_{(\text{para})}^\mu =- \frac{e^2}{2}\langle  \Psi^{\dagger}\check{v}^\mu \Psi\cdot\Psi^{\dagger}\left(\check{v}^\nu A^\nu\right)\Psi \rangle.
\end{align}
Therefore, the superfluid density acquires the contribution
\begin{align}
	\rho_{(\text{para})}
	=-\frac{\rmi}{2}
	\sumint_{\epsilon,\vec{p},\mu}\Tr\left[ \check{G}^K\check{v}^\mu \check{G}^R \check{v}^\mu
	+
	\check{G}^A\check{v}^\mu \check{G}^K \check{v}^\mu
	\right]
\end{align}
that is the sum of two qualitatively different terms. 
The details of the derivation are given in App.~\ref{sec:App_para}.
The first term
comes from quasiparticle intra-branch processes, and the ordering of
limits in the external frequency and in the momenta
$\omega,\vec{q}\rightarrow0$ is crucial. We focus on the case relevant
for the Meissner effect, \textit{i.e.} when the static limit (zero
frequency) is taken first, yielding
\begin{align}
\label{eq:rho_para_intra}
\rho_{(\text{para})}^{\text{intra}}
=
\sumint_{\vec{p},\mu} \!\! (V_\vec{p}^\mu+v_\vec{p}^\mu \cos\beta_\vec{p})^2 \underline{n}_{1\vec{p}}^\prime+(V_\vec{p}^\mu -v_\vec{p}^\mu \cos\beta_\vec{p})^2 \underline{n}_{2\vec{p}}^\prime,
\end{align}
where we introduced the quantities
\begin{align}
	\label{eq:n_derivative}
	\underline{n}_{1 (2) \vec{p}}^\prime\equiv\lim\limits_{\vec{q}\rightarrow0}\frac{\underline{n}_{1 (2) \vec{p}+\vec{q}}-\underline{n}_{ 1 (2) \vec{p}}}{(\varepsilon_{\vec{p} +\vec{q}} \pm \xi_{\vec{p} + \vec{q}}) - (\varepsilon_\vec{p} \pm \xi_\vec{p})}=\frac{\partial_p \underline{n}_{1 (2) p}}{v_{1 (2) p}^{\text{qp}}},
\end{align}
In the last step above, we assumed isotropic dispersion relations and $v_{\alpha}^{\text{qp}}$ is the quasiparticle velocity ($\alpha = 1, 2$).

The second component corresponds to quasiparticle inter-branch processes with pair creation and annihilation. Here, the order of limits $\omega,\vec{q}\rightarrow0$ does not matter. A straightforward calculation gives
\begin{align}
\label{eq:rho_para_inter}
\rho_{(\text{para})}^{\text{inter}}
= \frac{1}{2}
\int_\vec{p} v_\vec{p}^2\sin^2\beta_\vec{p} \, \frac{\underline{n}_{1 \vec{p}}+\underline{n}_{ 2 \vec{p}}-1}{\xi_\vec{p}}.
\end{align}

\subsection{Total superfluid density}
Integrating by parts the diamagnetic contribution in Eq.~(\ref{dia})
and summing with both paramagnetic terms, we obtain the final result
for the net superfluid density of the system (see appendix for
detailed steps):
\begin{align}
\label{rho}
\rho=-\int_{\vec{p}}(V_\vec{p}^2-v_\vec{p}^2)\sin^2\beta_\vec{p} \left[\frac{\underline{n}_{ 1 \vec{p}}+\underline{n}_{ 2 \vec{p}}-1}{\xi_\vec{p}} - \underline{n}_{ 1 \vec{p}}^\prime-\underline{n}_{ 2 \vec{p}}^\prime\right].
\end{align}
The equation above is the central result of our manuscript. It is
applicable to a wide class of electronic dispersions as long as the
superconducting state is stable and with a diagonal quasiparticle
distribution function. The immediate application of interest is to use
this result to study the superfluid density of a photo-induced
inter-band superconductor, which is the topic of the next Section.

\section{Superfluid density in a steady state}
In the optical-pumping setup presented in
Ref.~\onlinecite{goldstein2015}, a nonequilibrium population of the
two electron bands is created by shining a laser on the system. The
laser is responsible for the emergence of a resonance surface in
momentum space, $\S$, where the sum of the two bands $2 E_\vec{k} =
E_{1 \vec k}+E_{2 \vec{k}} = 0$. The resonant surface $\S$ has
essentially the same role as a Fermi surface in a conventional BCS
superconductor in thermal equilibrium. Superconducting pairing takes
place around the resonant surface $\S$, where $E_\vec{k}=0$.  For
simplicity, we assume that $\S$ is rotationally invariant. The
electronic dispersions can be expanded in the vicinity of this surface
as
\begin{align}
	E_k \equiv\frac{E_{1 k}+E_{2 k}}{2} \simeq V(k-k_\S)&+\kappa_+(k-k_\S)^2,
	\\
	\varepsilon_k \equiv\frac{E_{1 k}-E_{2 k}}{2} \simeq \varepsilon_0+v(k-k_\S)&+\kappa_-(k-k_\S)^2.
\end{align}
A situation which is especially favorable for the formation  of the superconducting order corresponds to whenever the velocity matching condition $V=0$ is satisfied, see Ref.~\onlinecite{goldstein2015} for details. Below, we concentrate on this case, such that $E_k\simeq\kappa_+(k-k_\S)^2$.

An optically-pumped system possesses a non-thermal distribution function. While, generically, the distribution can be parametrized as
\begin{align}
\Braket{\Psi_\vec{k}^{i\dagger}\Psi_\vec{k}^j}=
\begin{pmatrix}
n_{\vec{k}}^{11} & s_{\vec{k}}^{21}
\\
s_{\vec{k}}^{21*} & 1-n_{\vec{k}}^{22}
\end{pmatrix}\,,
\end{align}
in our specific case of photo-induced superconductivity there is a relation between the components of the above matrix\cite{goldstein2015},
\begin{align}
(2E+\rmi\Gamma_{12})s_{\vec{k}}^{21}=-\Delta^*(n_{\vec{k}}^{11}+n_{\vec{k}}^{22}-1),
\end{align}
where $\Gamma_{12}$ is an interband relaxation rate. At energies $E\gg\Gamma_{12}$, this relation implies a distribution function which is diagonal in the  quasiparticle basis and 
\begin{align} \label{eq:diag_n}
\Braket{\Psi_\vec{k}^{\alpha \dagger}\Psi_\vec{k}^\beta}
=
\hat{U}^\dagger\begin{pmatrix}
\underline{n}_{ 1 \vec{k}} & 0
\\
0 & 1 - \underline{n}_{2 \vec{k}}
\end{pmatrix}\hat{U} \;.
\end{align}
This is of course reasonable since quasiparticles are true excitations of the emergent superconducting state.

As it follows from Ref.\cite{goldstein2015}, the distribution functions in Eq.~(\ref{eq:diag_n}) are relatively smooth around the resonant surface and depend only on the energy $E_k$. Together with the assumed velocity matching condition,  $V\simeq 0$, this implies that the second term in the superfluid density in Eq.~(\ref{rho}) is negligible as compared to the first one, since the quasiparticle velocities $v_{\underline{1}(\underline{2}) p}=v_p \pm V_p \cos\beta_p \simeq v$ and
\begin{align}
\underline{n}_{\alpha p}^\prime \equiv 
\frac{\partial_k \underline{n}_{\alpha k}}{v_{\alpha k}^{\text{qp}}}\simeq\frac{d \underline{n}_{ \alpha}(E)} {dE} \, \left(\frac{V}{v}\right)\underset{V/v\rightarrow0}{\longrightarrow}0.
\end{align}
Simplifying, we get
\begin{align}
\rho\simeq\int_{\vec{p}} v^2\frac{\Delta^2}{\xi^3_p}(\underline{n}_{ 1 p}+\underline{n}_{ 2 p}-1)\,,
\end{align}
where, we recall, $\xi_\vec{p} \equiv \sqrt{E_\vec{p}^2 + \Delta^2} $.

\begin{table}[t]
	\begin{center}
		\begin{TAB}(@,25pt,25pt){|c|c|c|c|c|}{|c|c|c|c|c|}
			Superconductivity  
			& $\checkmark$ & $\checkmark$ & $\checkmark$ & $\checkmark$ 
			\\
			$V_{\rm int}$ 
			&- & - & +	& +  
			\\
			$N_\S$
			& - & + & -	& + 
			\\ 
			$\kappa_+ = \frac{\kappa_1 + \kappa_2}{2}$ 
			& + & - & -	& +  
			\\
			$\rho$ 
			& - & - & + & + 
		\end{TAB}
	\end{center}
	\caption{Sign of the superfluid density $\rho$ for different parameters: electronic interaction $V_{\rm int}$ (repulsive when positive), nonequilibrium population imbalance $N_\S$ [see Eq.~(\ref{eq:NF})] and average band curvature $\kappa_+$. Only the cases allowing for an inter-band superconducting state are displayed.}
	\label{table}
\end{table}

Now, we make use of the explicit form of distribution functions
obtained in Ref.~\onlinecite{goldstein2015} (see also the appendix
where we reproduce the derivation of the relevant results for the case
of quasiparticles in the steady state with a mean-field pairing field
and an external optical pump). In particular, a relevant quantity of
interest is
\begin{align}
	\underline{n}_{1\vec{k}}+ \underline{n}_{ 2 \vec{k}}-1=\frac{4 E_k \sqrt{E_k^2+\Delta^2}}{4E_k^2+\gamma_*\Delta^2} N_\S \, ,
\end{align}
where we introduced the combination
\begin{align} \label{eq:NF}
N_\S \equiv n_{\text{F}}(E_{1\vec{k}},\mu_1)+n_{\text{F}}(E_{2\vec{k}},\mu_2)-1 \,,
\end{align}
with $n_{\text{F}}(E_{\alpha\vec{k}},\mu_\alpha)$ the Fermi-Dirac
distributions corresponding to the quasi-thermal equilibrium that sets
up in each band. The effective chemical potentials $\mu_\alpha$ can be
seen as Lagrange multipliers enforcing the average number of particles
in each band, and depend on the balance between the optical drive and
the interband relaxation mechanisms (see
Ref.~\onlinecite{goldstein2015} for details). Note that the finiteness
of the above quantity, $N_\S \neq 0$, is crucial to the formation of a
interband Cooper pairing. For convenience we introduced $\gamma_*
\equiv \Gamma_{12}(\Gamma_1^{-1}+\Gamma_2^{-1})$ with $\Gamma_{1,2}$
being intraband relaxation rates. We obtain the superfluid density
\begin{align} \label{eq:rho137}
	\rho\simeq N_\S \,  v^2\int_{\vec{k}}\frac{4\Delta^2E_\vec{k}}{(E_\vec{k}^2+\Delta^2)(4E_\vec{k}^2+\gamma_*\Delta^2)} \,,
\end{align}
and  the self-consistency equation for photo-induced superconductivity
\begin{align} \label{eq:sc137}
	1=2 N_\S \, V_{\rm int}\int_{\vec{k}}\frac{E_\vec{k}}{4E_\vec{k}^2+\gamma_*\Delta^2} \,.
\end{align}
The most pressing questions are now 
\begin{itemize}
	\item Is superconductivity possible? (This question is the
          focus of Ref.~\onlinecite{goldstein2015}.)
	\item What is the sign of the superfluid density? (This is our
          main focus.)
\end{itemize}
Answers to these question depend solely on the signs of three parameters, namely: electron-electron interaction $V_{\rm int}$, curvature of the electron dispersion $\kappa_+$, and $N_\S$. 
Indeed, a rapid inspection of the right-hand side of Eq.~(\ref{eq:rho137}) implies that the sign of the superfluid density is governed by
\begin{align}
{\rm sgn}(\rho) = {\rm sgn}(\kappa_+) \times {\rm sgn}(N_\S)  \,.
\end{align}
Similarly, the inspection of the right-hand side of Eq.~(\ref{eq:sc137}) implies that a solution with a finite superconducting order parameter exists whenever
\begin{align}
{\rm sgn}(V_{\rm int}) = {\rm sgn}(\kappa_+) \times {\rm sgn}(N_\S)  \,.
\end{align}
The corresponding outcomes for different cases are summarized in the Table~\ref{table}.
We now compute the expression of the superfluid density in two limiting cases: $\Delta \gg \Gamma$ and $\Delta \ll \Gamma$.  

In the regime $\Delta \gg \Gamma$, we start by expanding the energy $E_k$ around the resonance surface (using the velocity matching condition, $V \simeq 0$), $E \simeq \kappa_+(k-k_\S)^2$, so that the superfluid density becomes
\begin{align}
	\label{rho2}
	\rho=(\sgn \kappa_+N_\S)\frac{A_\S v^2}{\sqrt{\Delta\abs{\kappa_+}}}B_\rho\left(\gamma_*\right)  \,,
\end{align}
where $A_\S=4\pi k_\S^2$ is the area of the resonant surface ($2\pi k_\S$ for a two-dimensional system) and $B_\rho(x)$ is a positive function
\begin{align}
	B_\rho(x)=\int_{-\infty}^\infty\d t\frac{4t^2}{(t^4+1)(4t^4+x)}
	=
	\begin{cases}
	\pi x^{-1/4} & x\ll1
	\\
	2\sqrt{2}\pi x^{-1} & x\gg1
	\end{cases} \,.
\end{align}
%
%
Equation~(\ref{rho2}) displays an anomalous scaling of the superfluid density $\rho$ with the order parameter $\Delta$,
\begin{align} \label{eq:scal1}
\rho \propto \frac{1}{\sqrt{\Delta} } \mbox{ for } \Delta \gg \Gamma \,.
\end{align}
Such a divergent scaling survives only while $\Delta\gg\Gamma$. In the regime where $\Delta\ll\Gamma$, after re-including properly the factors of $\Gamma$ that have been neglected so far (see the appendix for a detailed derivation), one finds 
\begin{align} \label{eq:scal2}
\rho\propto\Delta^2 \mbox{ for } \Delta \ll \Gamma \,,
\end{align}
which is similar to the conventional BCS scenario in equilibrium. The two scalings  in Eqs.~(\ref{eq:scal1}) and (\ref{eq:scal2}) signal that the superfluid density reaches a maximum in the crossover regime.

\section{Conclusion}

We have computed the superfluid density of the superconducting state
that can be induced by optically pumping valence band electrons to the
conduction band. We found a positive superfluid density in the
presence of repulsive electronic interactions, and this constitutes an
important check of the stability of the superconducting order
announced in Ref.~\onlinecite{goldstein2015}.  The next check is to
make sure that the heating caused by the optical pumping is slow
enough to allow for the superconducting order to develop (on the order
of hundreds on $1/\Gamma$'s) and, perhaps more importantly, for the
transport measurements to be performed.  The power dissipated can
estimated to be $\mathcal{P} \sim \Gamma_{\mathrm{Interband}} \times N
\times \hbar \omega_{\mathrm{gap}} $ with an interband recombination
rate $\Gamma_{\mathrm{Interband}} \sim 10^{-8}~e$V, a density of
states $N \sim 10^{20}$/cm$^3$, and $ \hbar \omega_{\mathrm{gap}} \sim
0.3~e$V, amounting to $\mathcal{P} \sim 10^{6-7}$~J/s.cm$^3$, and
leading to a generous window of $10^{-4}$~s to perform the experiments
before the sample temperature increases by approximately 10 K
(in the absence of external cooling).

\acknowledgments

This work has been supported by the DOE Grant DE-FG02-06ER46316
(C.C.), and by the Engineering and Physical Sciences Research Council
(EPSRC) and No. EP/M007065/1 (G.G.), and by the EPSRC Network Plus on
``Emergence and Physics far from Equilibrium''. Statement of
compliance with the EPSRC policy framework on research data: this
publication reports theoretical work that does not require supporting
research data.

\appendix
\section{Equations of motion for the quasiparticle distribution functions}
\label{appendix1}

\subsection{Equations of motion}

The equations of motion for the optical pumping setup employed in the
present paper are derived in Ref.~\onlinecite{goldstein2015}. They
read as:
\begin{equation}
\label{app:EoM}
\begin{split}
\frac{\d}{\d t}n^{11}_{\vec{k}} &= \rmi \Delta s^{12}_{\vec{k}}-\rmi \Delta^* s^{12*}_{\vec{k}}-2\Gamma_1\tilde{n}^{1}_{\vec{k}},
\\
\frac{\d}{\d t}n^{22}_{-\vec{k}} &= \rmi \Delta s^{12}_{\vec{k}}-\rmi \Delta^* s^{12*}_{\vec{k}}-2\Gamma_2\tilde{n}^{2}_{-\vec{k}},
\\
\frac{\d}{\d t}s^{12}_{\vec{k}} &= -\rmi (2E_{\vec{k}}-\rmi\Gamma_{12})s^{12}_{\vec{k}}
+\rmi\Delta^*(n^{11}_{\vec{k}}+n^{22}_{-\vec{k}}-1),
\end{split}
\end{equation}
where $n^{\alpha\alpha}_{\vec{k}}=\Braket{c^{\alpha\dagger}_{\vec{k}}c^{\alpha}_{\vec{k}}}$, and $s^{12}_{\vec{k}}=\Braket{c^{1\dagger}_{\vec{k}}c^{2\dagger}_{-\vec{k}}}$. The tilded quantities $\tilde{n}^{\alpha}_{\vec{k}}=n^{\alpha\alpha}_{\vec{k}}-n_{\text{F}}(E_{\alpha\vec{k}},\mu_\alpha)$, where $n_{\text{F}}(E_{\alpha\vec{k}},\mu_\alpha)$ are distribution functions in the external thermal bath (see Ref.~\onlinecite{goldstein2015} for details).

In the steady state all time derivatives are zero. The last equation gives a useful relation
\begin{equation}
	s^{12}_{\vec{k}}=\frac{\Delta^*}{2E_{\vec{k}}-\rmi \Gamma_{12}}(n^{11}_{\vec{k}}+n^{22}_{-\vec{k}}-1).
\end{equation}
Proceeding with solving equations of motion, we get
\begin{align}
	\label{app:nn}
	n^{11}_{\vec{k}}+n^{22}_{-\vec{k}}-1 = \frac{4E^2+\Gamma_{12}^2}{4E^2+\Gamma_{12}^2+\gamma_*\Delta^2}N_\S,
	\\
	\label{app:s12}
	s^{12}_{\vec{k}}=\frac{\Delta^*(2E+\rmi\Gamma_{12})}{4E^2+\Gamma_{12}^2+\gamma_*\Delta^2}N_\S
\end{align}
where $N_\S=(n_{\text{F}}(E_{1\vec{k}},\mu_1)+n_{\text{F}}(E_{2\vec{k}},\mu_2)-1)$ and $\gamma_*=\Gamma_{12}(\Gamma_1^{-1}+\Gamma_2^{-1})$. Only these two quantities, (\ref{app:nn}) and (\ref{app:s12}) are of interest, as we will see below.

\subsection{Quasiparticle distribution functions}

As we have stated in the main text, for energies larger than the decay
rates $\Gamma$ we deal with well-defined quasiparticle distribution
functions. Using equation (\ref{eq:n}), we get for the matrix
quasiparticle distribution function
\begin{equation}
\label{app:nQP}
\begin{split}
	\begin{pmatrix}
	n_{\underline{1}\vec{k}} & 0
	\\
	0 & 1-n_{\underline{2}\vec{k}}
	\end{pmatrix}
	=
	\hat{U}
	\begin{pmatrix}
	n^{11}_{\vec{k}} & s^{12}_{\vec{k}}
	\\
	s^{12*}_{\vec{k}} & 1-n^{22}_{-\vec{k}}
	\end{pmatrix}
	\hat{U}^\dagger.
\end{split}
\end{equation}

\begin{equation}
\begin{split}
	n_{\underline{1}\vec{k}}=n^{11}_{\vec{k}}\cos^2\frac{\beta}{2}
	+(1-n^{22}_{-\vec{k}})\sin^2\frac{\beta}{2}
	+(\mathrm{Re}\,s^{12}_{\vec{k}})\sin\beta,
	\\
	1-n_{\underline{2}\vec{k}}=n^{11}_{\vec{k}}\sin^2\frac{\beta}{2}
	+(1-n^{22}_{-\vec{k}})\cos^2\frac{\beta}{2}
	-(\mathrm{Re}\,s^{12}_{\vec{k}})\sin\beta.	
\end{split}
\end{equation}

As wee see in the main text, both the self consistency equation and the final approximation for the superfluid density depend only on the combination
\begin{equation}
	n_{\underline{1}\vec{k}}+n_{\underline{2}\vec{k}}-1
	=(n^{11}_{\vec{k}}+n^{22}_{-\vec{k}}-1)\cos\beta+2(\mathrm{Re}\,s^{12}_{\vec{k}})\sin\beta
\end{equation}

Recalling that we set the order parameter to be real and focusing on
energies larger than the decay rate,
\begin{equation}
n_{\underline{1}\vec{k}}+n_{\underline{2}\vec{k}}-1
=\frac{4EE_\Delta}{4E^2+\gamma_*\Delta^2}N_\S.
\end{equation}

\subsection{Offdiagonal element of the distribution function}

Finally, throughout the text we assumed that we deal with a pure quasiparticle state. The general form of the distribution function in the quasiparticle basis,

\begin{equation}
\begin{split}
\begin{pmatrix}
n_{\underline{1}\vec{k}} & \text{OD}
\\
\text{OD}^* & 1-n_{\underline{2}\vec{k}}
\end{pmatrix}
=
\hat{U}
\begin{pmatrix}
n^{11}_{\vec{k}} & s^{12}_{\vec{k}}
\\
s^{12*}_{\vec{k}} & 1-n^{22}_{-\vec{k}}
\end{pmatrix}
\hat{U}^\dagger,
\end{split}
\end{equation}
have an offdiagonal element OD. We implied that $\text{OD}=0$. Here we show that this statement is true. Using the definition of OD above, we get
\begin{equation}
	\text{OD}=\frac{1-n^{22}_{-\vec{k}}-n^{11}_{\vec{k}}}{2}\sin\beta
	+s_{12}\cos^2\frac{\beta}{2}-s_{12}^*\sin^2\frac{\beta}{2}.
\end{equation}
Making use of the equations (\ref{app:nn}) and (\ref{app:s12}),
\begin{equation}
\begin{split}
	\text{OD}
	=
	\frac{1}{4E^2+\Gamma_{12}^2+\gamma_*\Delta^2}
	\frac{N_\S}{2E_\Delta}\left(-\Delta(4E^2+\Gamma_{12}^2) + \right.
	\\
	\left.
	+4E^2\Delta+\rmi\Gamma_{12}\Delta
	\right),
\end{split}
\end{equation}
so that the offdiagonal element is indeed vanishingly small for
energies larger than decay rate $\Gamma_{12}$,
\begin{equation}
\begin{split}
\text{OD}
=
\frac{-\Delta\Gamma_{12}^2+\rmi\Gamma_{12}\Delta E_\Delta}{2E_\Delta(4E^2+\Gamma_{12}^2+\gamma_*\Delta^2)}
N_\S\underset{\Gamma_{12}\rightarrow0}{\rightarrow}0.
\end{split}
\end{equation}

\section{Paramagnetic contribution to the superfluid density} \label{sec:App_para}

Here we provide the details on the calculation of the Green function
bubble in the paramegnetic contribution to the density of states. As
we have mentioned in the main text, we have to be carefull with the
order of limits $\omega,\vec{q}\rightarrow0$ for the external vector
potential $\vec{A}$. The Meissner effect corresponds to expulsion of
the static magnetic field, so in order to get the superfluid density
we have to take the $\omega\rightarrow0$ limit first. To show the
importance of the order of limits we retain $\omega,\vec{q}$
explicitly:
\begin{align}
\rho_{(\text{para})}(\omega,\vec{q})
&=
\\
\nonumber
&=-\frac{\rmi}{2}
\sumint_{\epsilon,\vec{p},\mu}\Tr\left[ \check{G}^K_+\check{v}^\mu \check{G}^R_- \check{v}^\mu
+
\check{G}^A_+\check{v}^\mu \check{G}^K_- \check{v}^\mu
\right],
\end{align}
where $\pm$ correspond to arguments $\epsilon\pm\omega/2$,
$\vec{p}\pm\vec{q}/2$.

We separate the total Green's function bubble into intra- and interband parts based on whether the quasiparticles change bands within the Green's function bubble,
\begin{align}
\rho_{(\text{para})}(\omega,\vec{q})
=\rho_{(\text{para})}^{(\text{intra})}(\omega,\vec{q})+\rho_{(\text{para})}^{(\text{inter})}(\omega,\vec{q}).
\end{align}

\subsection{Intraband contribution}

The intraband contribution can in turn be broken down into contributions of the two quasiparticle bands,
\begin{align}
\rho_{(\text{para})}^{(\text{intra})}(\omega,\vec{q})
=\rho_{(\text{para})}^{(\text{intra-1})}(\omega,\vec{q})+\rho_{(\text{para})}^{(\text{intra-2})}(\omega,\vec{q}),
\end{align}
where the contribution of the first quasiparticle band is
\begin{align}
\label{app:para_intra_1}
\begin{split}
&\rho_{(\text{para})}^{(\text{intra-1})}(\omega,\vec{q})
=
\\
&=
-\frac{\rmi}{2}
\sumint_{\epsilon,\vec{p},\mu}(\ucheck{v}_{11}^\mu)^2\left( \underline{G}^K_{11+} \underline{G}^R_{11-}
+
\underline{G}^A_{11+} \underline{G}^K_{11-} 
\right),
\end{split}
\end{align}
The first term of (\ref{app:para_intra_1}) is
\begin{align}
&\sumint_{\epsilon,\vec{p},\mu} (\ucheck{v}_{11}^\mu)^2G^K_{1+} G^R_{1-}=
\\
\nonumber
&=
\sumint_{\epsilon,\vec{p},\mu} (\ucheck{v}_{11}^\mu)^2F_{11+}\left(G^R_{1+}-G^A_{1+}\right) G^R_{1-}
\\
\nonumber
&=
-2\rmi\pi\sumint_{\epsilon,\vec{p},\mu}(\ucheck{v}_{11}^\mu)^2 F_{11+}\delta(\epsilon_+-(\varepsilon+\xi)_+)\frac{1}{\epsilon_--(\varepsilon+\xi)_- + \rmi 0}
\\
\nonumber
&=-\rmi\sumint_{\vec{p},\mu}(\ucheck{v}_{11}^\mu)^2 \frac{F_{11+}}{(\varepsilon+\xi)_+-(\varepsilon+\xi)_--\omega + \rmi 0},
\end{align}
where we took into account the fact that the Green's function combination $\left(G^R-G^A\right)$ is related to density of states and
\begin{equation}
\begin{split}
G^R_{1+}-G^A_{1+}=&
\frac{1}{\epsilon_+-(\varepsilon+\xi)_+ + \rmi 0}-\frac{1}{\epsilon_+-(\varepsilon+\xi)_+ - \rmi 0}
\\
=&-2\rmi\pi\delta(\epsilon_+-(\varepsilon+\xi)_+),
\end{split}
\end{equation}
with $\delta$ being Dirac delta function.

Similarly, the second term turns out to be
\begin{align}
\begin{split}
&\sumint_{\epsilon,\vec{p},\mu} (\ucheck{v}_{11}^\mu)^2G^A_{1+} G^K_{1-}=
\\
&=
-\rmi\sumint_{\epsilon,\vec{p},\mu}(\ucheck{v}_{11}^\mu)^2 \frac{-F_{11-}}{(\varepsilon+\xi)_+-(\varepsilon+\xi)_--\omega + \rmi 0},
\end{split}
\end{align}
so that the total contribution of the first band into intraband part is
\begin{align}
\begin{split}
&\rho_{(\text{para})}^{(\text{intra-1})}(\omega,\vec{q})
=
\\
&=-\frac{1}{2}\sumint_{\epsilon,\vec{p},\mu}(\check{v}_{11}^\mu\check{v}_{11}^\mu) \frac{F_{11+}-F_{11-}}{(\varepsilon+\xi)_+-(\varepsilon+\xi)_--\omega + \rmi 0}.
\end{split}
\end{align}
To proceed we note that the matrix element $F_{11}$ gives the quasiparticle distribution function
\begin{equation}
	F_{11}(\varepsilon,\vec{p}) = 1-2\underline{n}_{1 \vec{p}},
\end{equation}
the matrix element of the velocity operator is
\begin{equation}
	\check{v}_{11}^\mu=V^\mu+v^\mu\cos\beta,
\end{equation}
and the limit yields the derivative (\ref{eq:n_derivative}) introduced in the main text
\begin{equation}
	\lim\limits_{\vec{q}\rightarrow0}\lim\limits_{\omega\rightarrow0}\frac{F_{11+}-F_{11-}}{(\varepsilon+\xi)_+-(\varepsilon+\xi)_--\omega + \rmi 0}=-2\underline{n}_{1\vec{p}}^\prime.
\end{equation}
Summing up, the contribution of the first quasiparticle band is
\begin{align}
\begin{split}
\rho_{(\text{para})}^{(\text{intra-1})}
=\sumint_{\vec{p},\mu}(V_{\vec{p}}^\mu+v_{\vec{p}}^\mu\cos\beta_{1\vec{p}})^2 \underline{n}_{1\vec{p}}^\prime.
\end{split}
\end{align}
Similarly, the contribution of the second band is
\begin{align}
\begin{split}
\rho_{(\text{para})}^{(\text{intra-2})}
=\sumint_{\vec{p},\mu}(V_{\vec{p}}^\mu-v_{\vec{p}}^\mu\cos\beta_{\vec{p}})^2 \underline{n}_{2\vec{p}}^\prime.
\end{split}
\end{align}
Summation of the two  reproduces the result (\ref{eq:rho_para_intra}) from the main text.

The order of limits in $\omega,\vec{q}\rightarrow0$ was crucial
throughout the calculation. With the opposite order of limits we would
have a zero intraband contribution
\begin{equation}
\begin{split}
&\lim\limits_{\omega\rightarrow0}\lim\limits_{\vec{q}\rightarrow0}\frac{\underline{n}_{1(\vec{p}+\vec{q}/2)}-\underline{n}_{1(\vec{p}-\vec{q}/2)}}{(\varepsilon+\xi)_{\vec{p}+\vec{q}/2}-(\varepsilon+\xi)_\vec{p}-\vec{q}/2-\omega + \rmi 0}=
\\
&=\lim\limits_{\omega\rightarrow0}\frac{0}{-\omega + \rmi 0}=0.
\end{split}
\end{equation}

Finally, we note that strictly speaking the vertex coupling electrons
to the vector potential is
\begin{equation}
	\frac{1}{2}\left(\check{v}_{\vec{p}+\vec{q}/2}^\mu+\check{v}_{\vec{p}-\vec{q}/2}^\mu\right)=\check{v}_{\vec{p}}^\mu+O(q^2),
\end{equation} 
but the corrections coming from finite external vector potential momentum $\vec{q}$  are irrelevant for the present calculation.

\subsection{Interband contribution}

In contrast, the order of limits is irrelevant for the interband
contribution due to the presence of the superconducting gap. In the
intraband contribution we compare energies of the two quasiparticles
from the same band and the quantity
$(\varepsilon+\xi)_+-(\varepsilon+\xi)_-\rightarrow0$ in the
denominator results in the importance of the order of
limits. Meanwhile, in the interband contribution below we will be
comparing two quasiparticles from different bands encountering a
well-defined denominator
$(\varepsilon+\xi)_+-(\varepsilon-\xi)_-\rightarrow2\xi\geq2\Delta$. Thus
we omit external frequency $\omega$ and momentum $\vec{q}$ right away.

Similarly to the intraband case, we can divide the interband
contribution into processes where the quasiparticle transitions from
the first band into the second $(1\rightarrow2)$, and in the opposite
direction $(2\rightarrow1)$,
\begin{align}
\rho_{(\text{para})}^{(\text{inter})}
=\rho_{(\text{para})}^{(\text{inter}(1\rightarrow2))}
+\rho_{(\text{para})}^{(\text{inter}(2\rightarrow1))}.
\end{align}
The contribution of $(1\rightarrow2)$ interband processes is
\begin{align}
\begin{split}
\rho_{(\text{para})}^{(\text{inter})}
=
-\frac{\rmi}{2}
\sumint_{\epsilon,\vec{p},\mu}(\ucheck{v}_{12}^\mu\ucheck{v}_{21}^\mu)&\left( \underline{G}^K_{11} \underline{G}^R_{22}
+
\underline{G}^A_{11} \underline{G}^K_{22}+\right.
\\
&\left.
+
\underline{G}^K_{22} \underline{G}^R_{11}
+
\underline{G}^A_{22} \underline{G}^K_{11} 
\right).
\end{split}
\end{align}
This expression can be conveniently represented as
\begin{align}
\begin{split}
\rho_{(\text{para})}^{(\text{inter})}
=
\Im
\sumint_{\epsilon,\vec{p},\mu}(\ucheck{v}_{12}^\mu\ucheck{v}_{21}^\mu)&\left( \underline{G}^K_{11} \underline{G}^R_{22}
+
\underline{G}^K_{22} \underline{G}^R_{11}
\right).
\end{split}
\end{align}
The first term in the equation above gives
\begin{align}
&\sumint_{\epsilon,\vec{p},\mu} (\ucheck{v}_{12}^\mu\ucheck{v}_{21}^\mu)\underline{G}^K_{11} \underline{G}^R_{22}=
\\
\nonumber
&=
\sumint_{\epsilon,\vec{p},\mu} (\ucheck{v}_{12}^\mu\ucheck{v}_{21}^\mu)F_{11}\left(\underline{G}^R_{11}-\underline{G}^A_{11}\right) \underline{G}^R_{22}
\\
\nonumber
&=
-2\rmi\pi\sumint_{\epsilon,\vec{p},\mu}(\ucheck{v}_{12}^\mu\ucheck{v}_{21}^\mu) F_{11}\delta(\epsilon-(\varepsilon+\xi))\frac{1}{\epsilon-(\varepsilon-\xi) + \rmi 0}
\\
\nonumber
&=-\rmi\sumint_{\vec{p},\mu}(\ucheck{v}_{12}^\mu\ucheck{v}_{21}^\mu) \frac{F_{11}}{2\xi}.
\end{align}
Similarly, the second term is
\begin{align}
\begin{split}
\sumint_{\epsilon,\vec{p},\mu} (\ucheck{v}_{12}^\mu\ucheck{v}_{21}^\mu)\underline{G}^K_{22} \underline{G}^R_{11}
=-\rmi\sumint_{\vec{p},\mu}(\ucheck{v}_{12}^\mu\ucheck{v}_{21}^\mu) \frac{F_{22}}{-2\xi}.
\end{split}
\end{align}
Taking into account that the diagonal elements of the quasiparticle distribution function are
\begin{equation}
\begin{split}
F_{11}(\varepsilon,\vec{p}) &= 1-2\underline{n}_{1 \vec{p}},
\\
F_{22}(\varepsilon,\vec{p}) &= 2\underline{n}_{2 \vec{p}}-1,
\end{split}
\end{equation}
and matrix elements of the velocity operator are
\begin{equation}
\check{v}_{12}^\mu=\check{v}_{21}^\mu=-v^\mu\sin\beta,
\end{equation}
we get
\begin{align}
\begin{split}
\rho_{(\text{para})}^{(\text{inter})}
=&
-
\sumint_{\vec{p},\mu}(v^\mu\sin\beta)^2\frac{F_{11}-F_{22}}{2\xi}
\\
=&
\int_{\vec{p},\mu}v_{\vec{p}}^2\sin^2\beta_{\vec{p}}\frac{\underline{n}_{1 \vec{p}}+\underline{n}_{2 \vec{p}}-1}{\xi_{\vec{p}}}
\end{split}
\end{align}
reproducing the result (\ref{eq:rho_para_inter}) from the main text.

\subsection{Total superfluid density}
Finally, while the summation of the dia- and paramagnetic contributions is a mathematical exercise, it is not entirely straightforward and we illustrate it here for the convenience of the reader. To reproduce the compact expression from the main text, we have to integrate by parts the diamagnetic contribution
\begin{align}
\begin{split}
\rho_{(\text{dia})}
=\sumint_{\vec{p},\mu} (\partial_\mu V^\mu_\vec{p} \cos\beta_\vec{p} \, (\underline{n}_{ 1 \vec{p}}+\underline{n}_{2 \vec{p}}-1)+
\\
+(\partial_\mu v^\mu_\vec{p})(\underline{n}_{ 1 \vec{p}}-\underline{n}_{ 2 \vec{p}}+1) \, .
\end{split}
\end{align}
Integrating by parts we have
\begin{align}
\begin{split}
\rho_{(\text{dia})}
=-\sumint_{\vec{p},\mu} ( V_\vec{p}^\mu \partial_\mu(\cos\beta_\vec{p}) \, (\underline{n}_{ 1 \vec{p}}+\underline{n}_{2 \vec{p}}-1)+
\\
+V_\vec{p}^\mu \cos\beta_\vec{p} \, \partial_\mu(\underline{n}_{ 1 \vec{p}}+\underline{n}_{2 \vec{p}})+
\\
+v_\vec{p}^\mu\partial_\mu(\underline{n}_{ 1 \vec{p}}-\underline{n}_{ 2 \vec{p}}) \, .
\end{split}
\end{align}
Let us label the contribution in each line above as $\rho_{(\text{dia})}^{(\alpha)}$, $\alpha=1,2,3$. 

First, since $\cos\beta_\vec{p}=E/\sqrt{E_\vec{p}^2+\Delta^2}$, we have
\begin{equation}
\partial_\mu(\cos\beta_\vec{p})=\frac{\Delta^2}{(E_\vec{p}^2+\Delta^2)^{3/2}}V_\vec{p}^\mu=\frac{V_\vec{p}^\mu}{\xi_\vec{p}}\sin^2\beta_\vec{p}
\end{equation}
and the first contribution becomes
\begin{align}
\rho_{(\text{dia})}^{(1)}
=-\int_{\vec{p},\mu}V_{\vec{p}}^2\sin^2\beta_{\vec{p}}\frac{\underline{n}_{1 \vec{p}}+\underline{n}_{2 \vec{p}}-1}{\xi_{\vec{p}}}.
\end{align}
Second, we go back to the definition of the derivative of the distribution function over the quasiparticle energy (\ref{eq:n_derivative})  and observe that since
\begin{equation}
	\underline{n}_{1(2)\vec{p}}^\prime\equiv\lim\limits_{\vec{q}\rightarrow0}\frac{\underline{n}_{1 (2) \vec{p}+\vec{q}}-\underline{n}_{ 1 (2) \vec{p}}}{(\varepsilon_{\vec{p} +\vec{q}} \pm \xi_{\vec{p} + \vec{q}}) - (\varepsilon_\vec{p} \pm \xi_\vec{p})},
\end{equation}
then
\begin{equation}
\begin{split}
	\partial_\mu\underline{n}_{1 (2) \vec{p}}=
	&\underline{n}_{1(2)\vec{p}}^\prime(v^\mu_\vec{p}\pm\partial^\mu\xi_\vec{p})
	\\
	=&\underline{n}_{1(2)\vec{p}}^\prime(v^\mu_\vec{p}\pm V_\vec{p}^\mu\cos\beta_\vec{p}).
\end{split}
\end{equation}
Using this identity, we have 
\begin{equation}
\begin{split}
	\rho_{(\text{dia})}^{(2)}
	=-\int_{\vec{p},\mu}V^\mu_\vec{p} \cos\beta_\vec{p} 
	\left(\underline{n}_{1\vec{p}}^\prime(v^\mu_\vec{p}+ V_\vec{p}^\mu\cos\beta_\vec{p})
	\right.
	\\
	\left.
	+\underline{n}_{2\vec{p}}^\prime(v^\mu_\vec{p}- V_\vec{p}^\mu\cos\beta_\vec{p})\right)
\end{split}
\end{equation}
and
\begin{equation}
\begin{split}
\rho_{(\text{dia})}^{(3)}
=-\int_{\vec{p},\mu}v_\vec{p}^\mu
\left(\underline{n}_{1\vec{p}}^\prime(v^\mu_\vec{p}+ V_\vec{p}^\mu\cos\beta_\vec{p})-
\right.
\\
\left.
-\underline{n}_{2\vec{p}}^\prime(v^\mu_\vec{p}- V_\vec{p}^\mu\cos\beta_\vec{p})\right),
\end{split}
\end{equation}
Adding together $\rho_{(\text{dia}-2)}$, $\rho_{(\text{dia}-3)}$ and $\rho_{(\text{para})}^{\text{intra}}$ (given by Eq.(\ref{eq:rho_para_intra})), we have
\begin{equation}
\begin{split}
&\rho_{(\text{dia})}^{(2)}+\rho_{(\text{dia})}^{(3)}+\rho_{(\text{para})}^{\text{intra}}
=
\\
&=-\int_{\vec{p},\mu}\left[
\underline{n}_{1\vec{p}}^\prime\left(v_\vec{p}^2+2v^\mu_\vec{p}V_\vec{p}^\mu\cos\beta_\vec{p}+V_\vec{p}^2\cos^2\beta_\vec{p}+
\right.\right.
\\
&\left.-(V_{\vec{p}}^\mu+v_{\vec{p}}^\mu\cos\beta_{1\vec{p}})^2\right)+
\\
&
+
\underline{n}_{2\vec{p}}^\prime\left(v_\vec{p}^2-2v^\mu_\vec{p}V_\vec{p}^\mu\cos\beta_\vec{p}+V_\vec{p}^2\cos^2\beta_\vec{p}-\right.
\\
&\left.\left.-(V_{\vec{p}}^\mu-v_{\vec{p}}^\mu\cos\beta_{1\vec{p}})^2\right)
\right].
\end{split}
\end{equation}
After simplification the expression becomes
\begin{equation}
\begin{split}
&\rho_{(\text{dia})}^{(2)}+\rho_{(\text{dia})}^{(3)}+\rho_{(\text{para})}^{\text{intra}}
=
\\
&=\int_{\vec{p},\mu}(V_\vec{p}^2-v_\vec{p}^2)\sin^2\beta_\vec{p}\left(\underline{n}_{1\vec{p}}^\prime+\underline{n}_{2\vec{p}}^\prime\right).
\end{split}
\end{equation}
The remaining piece $\rho_{(\text{dia}-1)}$ can be combined with the interband paramagnetic contribution $\rho_{(\text{para})}^{(\text{inter})}$
\begin{equation}
\begin{split}
&\rho_{(\text{dia})}^{(1)}+\rho_{(\text{para})}^{(\text{inter})}=
\\
&=-\int_{\vec{p},\mu}(V_\vec{p}^2-v_\vec{p}^2)\sin^2\beta_{\vec{p}}\frac{\underline{n}_{1 \vec{p}}+\underline{n}_{2 \vec{p}}-1}{\xi_{\vec{p}}}.
\end{split}
\end{equation}
It is clear now that we indeed obtain the total superfluid density from the main text:
\begin{align}
\rho=&\rho_{(\text{para})} + \rho_{(\text{dia})}
\\
\nonumber
=&\left(\rho_{(\text{dia})}^{(2)}+\rho_{(\text{dia})}^{(3)}+\rho_{(\text{para})}^{\text{intra}}\right)
+
\left(\rho_{(\text{dia})}^{(1)}+\rho_{(\text{para})}^{(\text{inter})}\right)
\\
\nonumber
=&
-\int_{\vec{p}}(V_\vec{p}^2-v_\vec{p}^2)\sin^2\beta_\vec{p} \left[\frac{\underline{n}_{ 1 \vec{p}}+\underline{n}_{ 2 \vec{p}}-1}{\xi_\vec{p}} - \underline{n}_{ 1 \vec{p}}^\prime-\underline{n}_{ 2 \vec{p}}^\prime\right].
\end{align}

\section{Finite $\Gamma$ effects on superfluid Density}

We would like to analyze more closely the superfluid density in the
vicinity of the superconducting transition. In order to explore the
region $\Delta\ll\Gamma$ we would like to phenomenologically include
the effects of finite dissipation $\Gamma$. Therefore we write:
\begin{equation}
\rho\cong N_\S v^{2}\int_{k}\frac{4\Delta^{2}E_{k}}{\left(E_{k}^{2}+\Gamma^{2}/4+\Delta^{2}\right)\left(4E_{k}^{2}+\Gamma^{2}+\gamma_{*}\Delta^{2}\right)}\label{eq:density_gamma}
\end{equation}

Furthermore the self consistency equation becomes
\begin{equation}
1=2 N_\S V_{\rm int}\int_{k}\frac{E_{k}}{\left(4E_{k}^{2}+\Gamma^{2}+\gamma_{*}\Delta^{2}\right)},
\end{equation}
which after the integration gives
\begin{equation}
1=\frac{\pi V_{\rm int}N_\S}{2\sqrt{2}\kappa_{+}}\times\left(\frac{\Gamma^{2}+\gamma_{*}\Delta^{2}}{4\kappa_{+}^{2}}\right)^{-1/4}\label{eq:Self_consistency_Gamma},
\end{equation}
From this expression we obtain that for superconductivity to exist we
must have that
\begin{equation}
V_{\rm int}>\frac{2\sqrt{\kappa_{+}\Gamma}}{\pi N_\S}\equiv V_{\rm min}.\label{eq:Minimal_coupling}
\end{equation}

We then get for $V_{\rm int}\geq V_{\rm min}$ and $\abs{V_{\rm int}-V_{\rm min}}\ll V_{\rm min}$
\begin{equation}
\begin{split}
\rho&\cong N_\S v^{2}\int_{k}\frac{4\Delta^{2}E_{k}}{\left(E_{k}^{2}+\Gamma^{2}/4\right)\left(4E_{k}^{2}+\Gamma^{2}\right)}  
\\
&=N_\S v^{2}\Delta^{2}\frac{1}{\kappa_{+}^{3}}\cdot\frac{\pi}{4\sqrt{2}}\left(\frac{\Gamma^{2}}{4\kappa_{+}^{2}}\right)^{-5/4}\label{eq:Superfluid_density}\\
& =v^{2}N_\S\Delta^{2}\frac{1}{\Gamma^{2}\sqrt{\kappa_{+}\Gamma}}\cdot\frac{\pi}{4\sqrt{2}}
\\
&\simeq\frac{\Delta^{2}}{2\sqrt{2}\Gamma^{2}}\frac{v^{2}}{V_{\rm min}},
\end{split}
\end{equation}
the result presented in the main text.

\vspace{3cm}

\bibliography{Meissner}

\begin{thebibliography}{13}%
\makeatletter
\providecommand \@ifxundefined [1]{%
 \@ifx{#1\undefined}
}%
\providecommand \@ifnum [1]{%
 \ifnum #1\expandafter \@firstoftwo
 \else \expandafter \@secondoftwo
 \fi
}%
\providecommand \@ifx [1]{%
 \ifx #1\expandafter \@firstoftwo
 \else \expandafter \@secondoftwo
 \fi
}%
\providecommand \natexlab [1]{#1}%
\providecommand \enquote  [1]{``#1''}%
\providecommand \bibnamefont  [1]{#1}%
\providecommand \bibfnamefont [1]{#1}%
\providecommand \citenamefont [1]{#1}%
\providecommand \href@noop [0]{\@secondoftwo}%
\providecommand \href [0]{\begingroup \@sanitize@url \@href}%
\providecommand \@href[1]{\@@startlink{#1}\@@href}%
\providecommand \@@href[1]{\endgroup#1\@@endlink}%
\providecommand \@sanitize@url [0]{\catcode `\\12\catcode `\$12\catcode
  `\&12\catcode `\#12\catcode `\^12\catcode `\_12\catcode `\%12\relax}%
\providecommand \@@startlink[1]{}%
\providecommand \@@endlink[0]{}%
\providecommand \url  [0]{\begingroup\@sanitize@url \@url }%
\providecommand \@url [1]{\endgroup\@href {#1}{\urlprefix }}%
\providecommand \urlprefix  [0]{URL }%
\providecommand \Eprint [0]{\href }%
\providecommand \doibase [0]{http://dx.doi.org/}%
\providecommand \selectlanguage [0]{\@gobble}%
\providecommand \bibinfo  [0]{\@secondoftwo}%
\providecommand \bibfield  [0]{\@secondoftwo}%
\providecommand \translation [1]{[#1]}%
\providecommand \BibitemOpen [0]{}%
\providecommand \bibitemStop [0]{}%
\providecommand \bibitemNoStop [0]{.\EOS\space}%
\providecommand \EOS [0]{\spacefactor3000\relax}%
\providecommand \BibitemShut  [1]{\csname bibitem#1\endcsname}%
\let\auto@bib@innerbib\@empty
\bibitem [{\citenamefont {Kaiser}\ \emph {et~al.}(2014)\citenamefont {Kaiser},
  \citenamefont {Hunt}, \citenamefont {Nicoletti}, \citenamefont {Hu},
  \citenamefont {Gierz}, \citenamefont {Liu}, \citenamefont {Le~Tacon},
  \citenamefont {Loew}, \citenamefont {Haug}, \citenamefont {Keimer},\ and\
  \citenamefont {Cavalleri}}]{kaiser2014}%
  \BibitemOpen
  \bibfield  {author} {\bibinfo {author} {\bibfnamefont {S.}~\bibnamefont
  {Kaiser}}, \bibinfo {author} {\bibfnamefont {C.~R.}\ \bibnamefont {Hunt}},
  \bibinfo {author} {\bibfnamefont {D.}~\bibnamefont {Nicoletti}}, \bibinfo
  {author} {\bibfnamefont {W.}~\bibnamefont {Hu}}, \bibinfo {author}
  {\bibfnamefont {I.}~\bibnamefont {Gierz}}, \bibinfo {author} {\bibfnamefont
  {H.~Y.}\ \bibnamefont {Liu}}, \bibinfo {author} {\bibfnamefont
  {M.}~\bibnamefont {Le~Tacon}}, \bibinfo {author} {\bibfnamefont
  {T.}~\bibnamefont {Loew}}, \bibinfo {author} {\bibfnamefont {D.}~\bibnamefont
  {Haug}}, \bibinfo {author} {\bibfnamefont {B.}~\bibnamefont {Keimer}}, \ and\
  \bibinfo {author} {\bibfnamefont {A.}~\bibnamefont {Cavalleri}},\ }\href
  {\doibase 10.1103/PhysRevB.89.184516} {\bibfield  {journal} {\bibinfo
  {journal} {Phys. Rev. B}\ }\textbf {\bibinfo {volume} {89}},\ \bibinfo
  {pages} {184516} (\bibinfo {year} {2014})}\BibitemShut {NoStop}%
\bibitem [{\citenamefont {Mankowsky}\ \emph {et~al.}(2016)\citenamefont
  {Mankowsky}, \citenamefont {FÃ¶rst},\ and\ \citenamefont
  {Cavalleri}}]{Review_Cavalleri}%
  \BibitemOpen
  \bibfield  {author} {\bibinfo {author} {\bibfnamefont {R.}~\bibnamefont
  {Mankowsky}}, \bibinfo {author} {\bibfnamefont {M.}~\bibnamefont {FÃ¶rst}},
  \ and\ \bibinfo {author} {\bibfnamefont {A.}~\bibnamefont {Cavalleri}},\
  }\href {http://stacks.iop.org/0034-4885/79/i=6/a=064503} {\bibfield
  {journal} {\bibinfo  {journal} {Reports on Progress in Physics}\ }\textbf
  {\bibinfo {volume} {79}},\ \bibinfo {pages} {064503} (\bibinfo {year}
  {2016})}\BibitemShut {NoStop}%
\bibitem [{\citenamefont {{Gor'kov}}\ and\ \citenamefont
  {{Eliashberg}}(1968)}]{gorkov1968}%
  \BibitemOpen
  \bibfield  {author} {\bibinfo {author} {\bibfnamefont {L.~P.}\ \bibnamefont
  {{Gor'kov}}}\ and\ \bibinfo {author} {\bibfnamefont {G.~M.}\ \bibnamefont
  {{Eliashberg}}},\ }\href@noop {} {\bibfield  {journal} {\bibinfo  {journal}
  {JETP Lett}\ }\textbf {\bibinfo {volume} {8}},\ \bibinfo {pages} {202}
  (\bibinfo {year} {1968})}\BibitemShut {NoStop}%
\bibitem [{\citenamefont {Eliashberg}(1970)}]{eliashberg1970}%
  \BibitemOpen
  \bibfield  {author} {\bibinfo {author} {\bibfnamefont {G.~M.}\ \bibnamefont
  {Eliashberg}},\ }\href@noop {} {\bibfield  {journal} {\bibinfo  {journal}
  {JETP Lett.}\ }\textbf {\bibinfo {volume} {11}},\ \bibinfo {pages} {114}
  (\bibinfo {year} {1970})}\BibitemShut {NoStop}%
\bibitem [{\citenamefont {Dmitriev}\ \emph {et~al.}(1970)\citenamefont
  {Dmitriev}, \citenamefont {Khristenko}, \citenamefont {Trubitsyn},\ and\
  \citenamefont {Mende}}]{dmitriev1970}%
  \BibitemOpen
  \bibfield  {author} {\bibinfo {author} {\bibfnamefont {V.~M.}\ \bibnamefont
  {Dmitriev}}, \bibinfo {author} {\bibfnamefont {E.~V.}\ \bibnamefont
  {Khristenko}}, \bibinfo {author} {\bibfnamefont {A.~V.}\ \bibnamefont
  {Trubitsyn}}, \ and\ \bibinfo {author} {\bibfnamefont {F.~F.}\ \bibnamefont
  {Mende}},\ }\href@noop {} {\bibfield  {journal} {\bibinfo  {journal} {Ukr.
  Fiz. Zh.}\ }\textbf {\bibinfo {volume} {15}},\ \bibinfo {pages} {1614}
  (\bibinfo {year} {1970})}\BibitemShut {NoStop}%
\bibitem [{\citenamefont {Galitskii}\ \emph {et~al.}(1973)\citenamefont
  {Galitskii}, \citenamefont {Elesin},\ and\ \citenamefont
  {Kopaev}}]{galitskii1973}%
  \BibitemOpen
  \bibfield  {author} {\bibinfo {author} {\bibfnamefont {V.}~\bibnamefont
  {Galitskii}}, \bibinfo {author} {\bibfnamefont {V.}~\bibnamefont {Elesin}}, \
  and\ \bibinfo {author} {\bibfnamefont {Y.~V.}\ \bibnamefont {Kopaev}},\
  }\href@noop {} {\bibfield  {journal} {\bibinfo  {journal} {JETP Lett}\
  }\textbf {\bibinfo {volume} {18}},\ \bibinfo {pages} {27} (\bibinfo {year}
  {1973})}\BibitemShut {NoStop}%
\bibitem [{\citenamefont {Kirzhnits}\ and\ \citenamefont
  {Kopaev}(1973)}]{kirzhnits1973}%
  \BibitemOpen
  \bibfield  {author} {\bibinfo {author} {\bibfnamefont {D.}~\bibnamefont
  {Kirzhnits}}\ and\ \bibinfo {author} {\bibfnamefont {Y.~V.}\ \bibnamefont
  {Kopaev}},\ }\href@noop {} {\bibfield  {journal} {\bibinfo  {journal} {JETP
  Lett}\ }\textbf {\bibinfo {volume} {17}},\ \bibinfo {pages} {270} (\bibinfo
  {year} {1973})}\BibitemShut {NoStop}%
\bibitem [{\citenamefont {Elesin}\ \emph {et~al.}(1974)\citenamefont {Elesin},
  \citenamefont {Kopaev},\ and\ \citenamefont {Timerov}}]{elesin1973theory}%
  \BibitemOpen
  \bibfield  {author} {\bibinfo {author} {\bibfnamefont {V.}~\bibnamefont
  {Elesin}}, \bibinfo {author} {\bibfnamefont {Y.~V.}\ \bibnamefont {Kopaev}},
  \ and\ \bibinfo {author} {\bibfnamefont {R.~K.}\ \bibnamefont {Timerov}},\
  }\href@noop {} {\bibfield  {journal} {\bibinfo  {journal} {JETP Lett}\
  }\textbf {\bibinfo {volume} {38}},\ \bibinfo {pages} {1170} (\bibinfo {year}
  {1974})}\BibitemShut {NoStop}%
\bibitem [{\citenamefont {Galitskii}\ \emph {et~al.}(1986)\citenamefont
  {Galitskii}, \citenamefont {Elesin},\ and\ \citenamefont
  {Kopaev}}]{galitskii1986}%
  \BibitemOpen
  \bibfield  {author} {\bibinfo {author} {\bibfnamefont {V.~M.}\ \bibnamefont
  {Galitskii}}, \bibinfo {author} {\bibfnamefont {V.~F.}\ \bibnamefont
  {Elesin}}, \ and\ \bibinfo {author} {\bibfnamefont {Y.~V.}\ \bibnamefont
  {Kopaev}},\ }in\ \href@noop {} {\emph {\bibinfo {booktitle} {Nonequilibrium
  Superconductivity}}},\ \bibinfo {editor} {edited by\ \bibinfo {editor}
  {\bibfnamefont {D.~N.}\ \bibnamefont {Langenberg}}\ and\ \bibinfo {editor}
  {\bibfnamefont {A.~I.}\ \bibnamefont {Larkin}}}\ (\bibinfo  {publisher}
  {Elsevier Science Publishers},\ \bibinfo {year} {1986})\ pp.\ \bibinfo
  {pages} {377--451}\BibitemShut {NoStop}%
\bibitem [{\citenamefont {Goldstein}\ \emph {et~al.}(2015)\citenamefont
  {Goldstein}, \citenamefont {Aron},\ and\ \citenamefont
  {Chamon}}]{goldstein2015}%
  \BibitemOpen
  \bibfield  {author} {\bibinfo {author} {\bibfnamefont {G.}~\bibnamefont
  {Goldstein}}, \bibinfo {author} {\bibfnamefont {C.}~\bibnamefont {Aron}}, \
  and\ \bibinfo {author} {\bibfnamefont {C.}~\bibnamefont {Chamon}},\ }\href
  {\doibase 10.1103/PhysRevB.91.054517} {\bibfield  {journal} {\bibinfo
  {journal} {Phys. Rev. B}\ }\textbf {\bibinfo {volume} {91}},\ \bibinfo
  {pages} {054517} (\bibinfo {year} {2015})}\BibitemShut {NoStop}%
\bibitem [{\citenamefont {Keldysh}(1965)}]{keldysh}%
  \BibitemOpen
  \bibfield  {author} {\bibinfo {author} {\bibfnamefont {L.~V.}\ \bibnamefont
  {Keldysh}},\ }\href@noop {} {\bibfield  {journal} {\bibinfo  {journal} {Sov.
  Phys. JETP}\ }\textbf {\bibinfo {volume} {20}},\ \bibinfo {pages} {1018}
  (\bibinfo {year} {1965})}\BibitemShut {NoStop}%
\bibitem [{\citenamefont {Rammer}\ and\ \citenamefont {Smith}(1986)}]{Rammer}%
  \BibitemOpen
  \bibfield  {author} {\bibinfo {author} {\bibfnamefont {J.}~\bibnamefont
  {Rammer}}\ and\ \bibinfo {author} {\bibfnamefont {H.}~\bibnamefont {Smith}},\
  }\href {\doibase 10.1103/RevModPhys.58.323} {\bibfield  {journal} {\bibinfo
  {journal} {Rev. Mod. Phys.}\ }\textbf {\bibinfo {volume} {58}},\ \bibinfo
  {pages} {323} (\bibinfo {year} {1986})}\BibitemShut {NoStop}%
\bibitem [{\citenamefont {Kamenev}\ and\ \citenamefont
  {Levchenko}(2009)}]{Kamenev2009}%
  \BibitemOpen
  \bibfield  {author} {\bibinfo {author} {\bibfnamefont {A.}~\bibnamefont
  {Kamenev}}\ and\ \bibinfo {author} {\bibfnamefont {A.}~\bibnamefont
  {Levchenko}},\ }\href@noop {} {\bibfield  {journal} {\bibinfo  {journal}
  {Advances in Physics}\ }\textbf {\bibinfo {volume} {58}},\ \bibinfo {pages}
  {197} (\bibinfo {year} {2009})}\BibitemShut {NoStop}%
\end{thebibliography}%

\end{document}